\documentstyle[prl,aps,epsfig,twocolumn]{revtex}

\begin{document}
\draft

\twocolumn[\hsize\textwidth\columnwidth\hsize\csname@twocolumnfalse\endcsname
\title{A lossy transmission line as a quantum open system in the standard
quantum limit }
\author{ Y. D. Wang and C. P. Sun$^{a,b}$ }
\address{Institute of Theoretical Physics, the Chinese Academy of Sciences,\\
Beijing, 100080, China}
\maketitle

\begin{abstract}
We systematically investigate how to quantize a transmission line
resonator (TLR) in a mesoscopic electrical circuit in the presence
of the resistance and the conductance of the dielectric media.
Developed from the quantum bath based effective Hamiltonian method
for single mode harmonic oscillator, the approach we presented in
this article is a microscopic theory integrating quantum
fluctuation-dissipation relation. To qualitatively check the
condition under which the TLR can behave as a quantum object we
study the classical-quantum boundary characterized by the standard
quantum limit.
\end{abstract}


]

An ideal one-dimensional transmission line resonator (TLR) can be
described by a classical wave equation and thus can be quantized
\cite{Louisell} as usual by following a standard procedure - the
canonical quantization approach \cite{dirac}. Such a quantization
formalism is not treated as a serious issue in the usual
realities. Most recently, however, the situation has been changed
radically due to the rapid progresses in solid state based quantum
computing (QC). In one of such QC schemes a one-dimensional TLR is
used to coherently couple one or more Josephson junction (JJ) qubits \cite%
{Yele1,Yele2,yele3,Europl}. In order to create controllable
quantum entanglements among these JJ qubits, the TLR has to work
in a quantum manner as a quantum data bus linking these qubits.
Otherwise the TLR can not induce an effective inter-qubit
interaction. One can imagine this in the conventional cavity QED:
the classical cavity mode in the strong field limit do not induce
inter-atom interactions to form quantum entanglement of the
qubits.

The above arguments show that the validity of the TLR based
quantum computing strongly depends on whether the mode of TLR is
truly quantized, that is, has some observable quantum effects. In
this sense, the way to quantize the modes of the TLR and the
corresponding quantization condition become fundamentally
important for applications of TLR in quantum information
processing. In this article we devote to answer this question in a
more realistic situation taking into consideration of leakage.
Furthermore, we will try to find out what characterize the
boundary between the quantum and classical regime for the lossy
TLR by examining the so-called standard quantum limit (SQL)
\cite{landua,Bra}, which, as a consequence of the Heisenberg
uncertainty principle, is usually referred to as the fundamental
limit of the precision of repeated position measurements. In the
modern technology based quantum measurement it was recognized that
if one can reach the SQL in the experiments, the quantum behavior
is observable even for macroscopic objects. Most recently LaHaye
\emph{et. al.}\cite{nano} described such an experiment with the
goal to test SQL on a vibrating nano-mechanical beam that is about
one-hundredth of a millimeter. This excites our interest on the
similar problem about the actual boundary between the classical
and quantum regime for the TLR.

To start with we consider the quantization of the modes of the
one-dimensional lossy TLR. In our model the lossy TLR is treated
as an open system interacting with a bath, a thermal environment.
It may be a background electromagnetic field interacting with the
transmitted charge in the TLR, or the classical lead connected to
the TLR, or other damping mechanism. Mathematically the lossy TLR
can be well depicted by a wave equation with leakage and its
Fourier components obey the typical dissipation equation. In this
sense, the idea and methods developed in our previous works
\cite{ys1,Sun,Gao} on quantum dissipative system can be applied to
the present discussion. We will show that the interaction with the
bath leads to an explicit description for the TLR in terms of the
generalized Caldirora-Kani (CK) effective Hamiltonian. We also
examine how the quantum and thermal fluctuations of the
environment contribute to the uncertainty of the canonical
variables of the lossy TLR.

To be universal we firstly revisit the derivation of the classical
wave function for the lossy TLR in high dimensional case. The
model is depicted by four lumped parameters, the distributed
resistance $r$, the distributed conductance $g$ of the dielectric
media, the distributed inductance $l$ and
distributed capacitance $c$ per unit length. Let $V=V({\bf x},t)$ and ${\bf %
I=I(x,}t{\bf )}$ be the distributions of the voltage and the
current vector respectively. With the conservation of the current,
the Kirchhoff's voltage law leads to the equations of motion
\cite{rem1}

\begin{equation}
\nabla V=-r{\bf I}-l\frac{\partial {\bf I}}{\partial t},\quad \nabla \cdot
{\bf I}=-gV-c\frac{\partial V}{\partial t}  \label{cc}
\end{equation}%
Eliminating the current vector in the above two equations we obtain the high
dimensional lossy wave equation for the voltage \cite{rma}.

\begin{equation}
\nabla ^{2}V=rgV+\left( rc+gl\right) \frac{\partial V}{\partial t}+\frac{1}{%
v^{2}}\frac{\partial ^{2}V}{\partial t^{2}}
\end{equation}%
where $v=\sqrt{\frac{1}{lc}}$ is the velocity of propagation or
phase speed. For an isolated conductor there does not exist the
current along the norm direction of the surface of conductor and
one thus have the boundary condition ${\bf I}_{{\bf
n}}|_{boundary}\equiv{\bf I\cdot n}|_{boundary}=0$ where ${\bf n}$
is the direction of the norm of the boundary surface. Combining
with the equations of motion, Eq.(1) leads to the boundary
condition ${\bf n\cdot }\nabla V|_{boundary}={\bf n\cdot }\left( -r{\bf I}-l%
\frac{\partial {\bf I}}{\partial t}\right) {\bf =}0$ for the
voltage equation. To be specific in one dimensional case, we have
the boundary conditions for the current $I(x=0,t)=I(x=L,t)=0$ of
an isolated 1-d TLR of
length $L$. The corresponding boundary conditions for voltage equation is $%
V_{x}(x=0,t)=V_{x}(x=L,t)=0$. In the following discussion we will
focus on the quantization of the 1-d lossy TLR with such boundary
conditions.

A canonical quantization scheme for an ideal 1-d lossless TLR can
be found in Louisell's monograph with an implied Hamiltonian
\cite{Louisell}. But for the lossy TLR as an open system the
energy does not conserve and thus there does not exist a
time-independent Hamiltonian a priori. Hence it is necessary to
develope a quantization scheme only based on the classical
equation of motion. In the very original paper of quantum
mechanics by Heisenberg, however, without knowing the Hamiltonian
or Lagrangian beforehand the classical equation of motion is
sufficient to solve the quantized energy levels of linear
oscillator. The modern version of this idea was presented in the
famous textbook by Landau and Lifshitz \cite{ll}.
Following this method we even quantized the quantum dissipation system \cite%
{ys1,Sun,Gao} to re-deduce the so-called CK effective Hamiltonian \cite%
{C,K,Dikker} directly from the classical equation of motion for the open
system with the linear coupling to the Ohmic bath \cite{legget}.

Now we quantize the modes of the lossy TLR by developing our previous
approach mentioned above. We can start from the lossy voltage equation
\begin{equation}
\frac{1}{v^{2}}\frac{\partial ^{2}}{\partial t^{2}}V+\left( lg+rc\right)
\frac{\partial }{\partial t}V+rgV-\frac{\partial ^{2}V}{\partial x^{2}}=0
\label{4}
\end{equation}%
with boundary condition $V_{x}\left( 0,t\right) =V_{x}\left(
L,t\right) =0$. The time-dependent Fourier's components
$V_{n}\left( t\right) $ of $V(x,t)$ corresponding the normal mode
$\sqrt{\frac{2}{L}}\cos \frac{n\pi x}{L}$ just satisfies the
dissipation equation
\begin{equation}
\frac{d^{2}}{dt^{2}}V_{n}(t)+\gamma \frac{d}{dt}V_{n}(t)+\omega
_{n}^{2}V_{n}(t)=0  \label{5}
\end{equation}
where the dissipation rate and the effective frequency are $%
\gamma =v^{2}(gl+rc)$ and $\omega _{n}^{2}=v^{2}[rg+\left( \frac{n\pi }{L}%
\right) ^{2}]$ respectively.

To find out the canonical commutation relation for the dynamical variables $%
V_{n}\left( t\right) $ and $\frac{d}{dt}V_{n}(t)$ we calculate the time
evolution of their commutator $B(t)=[V_{n},\frac{d}{dt}V_{n}\left( t\right) ]
$ using the equation of motion (\ref{5}), obtaining the close equation $%
\frac{d}{dt}B=-\gamma B$. That means $B\propto e^{-\gamma t}$, or
\begin{equation}
[V_{n},\frac{d}{dt}V_{n}\left( t\right) ]=i\hbar e^{-\gamma t}/M_{n}.
\label{8}
\end{equation}
Here we choose the effective mass $M_{n}=\frac{c}{\omega
_{n}^{2}}$ so that we can get a correct expression for the energy
of TLR. Therefore, we can define the canonical momentum operators
\begin{equation}
P_{n}\left( t\right) =M_{n}e^{\gamma t}\frac{d}{dt}V_{n}\left( t\right)
\end{equation}
to realize the conventional canonical commutation relation $\left[
V_{n},P_{m}\right] =i\hbar \delta _{nm}$.

In terms of the annihilation and creation operators $a_{n}\left( t\right) $
and $a_{n}^{\dag }\left( t\right) $ defined by
\begin{eqnarray}
V_{n}\left( t\right) &=&\sqrt{\frac{\hbar \omega _{n}}{2c}}\left(
a_{n}+a_{n}^{\dag }\right) ,  \nonumber \\
P_{n}\left( t\right) &=&-i\sqrt{\frac{\hbar c }{2\omega
_{n}}}\left( a_{n}-a_{n}^{\dag }\right)  \label{11}
\end{eqnarray}
the C-K Hamiltonian for the 1-D lossy TLR is obtained as ${\cal {H}}%
=\sum_{n}H_{n}$, where
\begin{equation}
H_{n}\equiv \hbar \omega _{n}\{[a_{n}^{\dag 2}\sinh (\gamma t)+a_{n}^{\dag
}a_{n}\cosh (\gamma t)]+h.c\}.
\end{equation}
Here we have ignored a time-dependent c-number $\hbar \omega
_{n}\cosh (\gamma t)$

The above effective Hamiltonian ${\cal H}$ can force the TLR to
evolve into the multi-mode squeezed state when the TLR is
initially prepared in a multi-mode coherent state $\left\vert
\alpha \right\rangle \equiv \left\vert \alpha _{1},\alpha
_{2,}...,\alpha _{n},...\right\rangle $ where $\left\vert \alpha
\right\rangle _{n}$ denotes the $n$-th mode. This conclusion can
be proved by rewriting $H_{n}$ as
\begin{equation}
H=\hbar \omega _{n}^{\prime }A_{n}^{\dag }A_{n}=S_{n}^{\dag }\left( t\right)
\hbar \omega _{n}a_{n}^{\dag }a_{n}S_{n}\left( t\right)
\end{equation}
Here, $\{A_{n}\}$ is a new set of bosonic operator defined as the
unitary transformation of $a_{n}$ $A_{n}=S_{n}^{\dag }\left(
t\right) a_{n}S_{n}\left( t\right) $ by the squeezing operators
\cite{WYD}
\begin{equation}
S_{n}\left( t\right) =\exp \left[ \frac{1}{4}\gamma t(a_{n}^{\dag
2}-a_{n}^{2})\right]
\end{equation}
We can also write down the explicit expressions $A_{n}=\xi a_{n}-\eta
a_{n}^{\dag }$ with $\xi =\cosh (\gamma t/2),\eta =-\sinh (\gamma t/2)$.
Then we obtain the evolution operator
\begin{equation}
U_{n}\left( t\right) =S_{n}^{\dag }\left( t\right) \exp [-i\omega
_{n}a_{n}^{\dag }a_{n}t]S_{n}\left( 0\right),
\end{equation}
which gives the the final state $\left\vert \Psi _{n}\left( t\right)
\right\rangle =U_{n}\left( t\right) \left\vert \alpha \right\rangle
_{n}=\left\vert \alpha e^{-i\omega _{n}t},\xi ,-\eta \right\rangle _{n}$ as
a squeezed state defined as the eigenstate of $A_{n}$ with eigenvalue $%
\alpha e^ {-i\omega _{n}t}$.

Studying the above results carefully, it seems that the above
results are not totally convincing due to the violation of the
uncertainty principle about coordinate $V_{n}\left( t\right) $ and
momentum $P_{n}\left( t\right)$ because the above arguments are
too phenomenological and the source of dissipation is not
considered microscopically. Thus the Brownian motion can not be
analyzed in the frame of the effective CK Hamiltonian formalism.
What's more, if we would use the above CK Hamiltonian without
restriction, some ridiculous conclusions are to be reached.
Fortunately, we can solve this problem by demonstrating the
derivation of the phenomenological CK Hamiltonian, starting with
the conventional system-plus-reservoir approach. Suppose the
environment is a bath of many harmonic oscillators linearly
coupled to the open system. The bath and system constitute a
conservative composite system and thus its quantization is rather
straightforward. As shown in our previous works, the total wave
function is partially factorized with respect to system and bath
when the Brownian fluctuation can be ignored under certain
conditions \cite{ys1}. the factorized part of the system is just
the CK effective wave function governed by the CK Hamiltonian.
With the these recalls we now consider the damping mode equation
\begin{equation}
\stackrel{\cdot \cdot }{V}_{n}(t)+\gamma \stackrel{\cdot }{V}_{n}(t)+\omega
_{n}^{2}V_{n}(t)=f_{n}\left( t\right)  \label{13}
\end{equation}%
where the parameter $\gamma $ and $\omega _{n}$ are the same as before and
\begin{equation}
f_{n}(t)=-\sum_{j}c_{jn}(x_{nj0}\,\cos \omega _{nj}t+\dot{x}_{nj0}\frac{\sin
\omega _{nj}t}{\omega _{nj}})
\end{equation}%
is a Brownian driving force. Here, $x_{j0}$ ($\,\dot{x}_{j0}$) are
the initial values of the canonical coordinate (its first
derivative with respect to $t$) of the harmonic oscillators of the
bath that coupled to the transmission line, $\omega _{nj}$ is the
frequency of the $j$-th oscillator of the independent reservoir
coupling to the $n$-th mode of QTL with the corresponding coupling
constant $c_{jn}$.

The solution $V_{n}\left( t\right) \equiv Q_{n}\left( t\right)
+\sum_{j}\xi _{nj}\left( t\right) $ of the above motion equation
(\ref{13}) can be solved as a direct sum of the dissipative motion
\begin{equation}
Q_{n}\left( t\right) =a_{1}\left( t\right) V_{n0}+a_{2}\left( t\right) \dot{V%
}_{n0}
\end{equation}%
and the quantum fluctuation of the reservoir.
\begin{equation}
\xi _{nj}\left( t\right) =\sum_{j}b_{nj1}\left( t\right)
x_{nj0}+b_{nj2}\left( t\right) \dot{x}_{nj0}
\end{equation}%
where the coefficients $a_{1}\left( t\right) =\frac{1}{\nu -\mu }(\nu
e^{-\mu t}-\mu e^{-\nu t})$, $a_{2}\left( t\right) =\frac{1}{\nu -\mu }%
(e^{-\mu t}-e^{-\nu t})$ with $\mu =\frac{\gamma }{2}+i\omega _{n}^{\prime }$%
, $\nu =\frac{\gamma }{2}-i\omega _{n}^{\prime }$, $\omega _{n}^{\prime }=%
\sqrt{\omega _{n}^{2}-\frac{\gamma ^{2}}{4}}$ and we ignore the lengthy and
too tedious expressions of $b_{nj1}( t) $ and $b_{nj2}\left( t\right) .$

To revisit the quantum effect of QTL, we need to consider the
standard quantum limit (SQL). If the measurement that probes it
can reach the accuracy of the SQL, then the quantum effect is
observable. Obviously the standard quantum limit of $V_{n}$ is
contributed by the measurement of QTL and the bath fluctuation
\begin{equation}
(\Delta V_{n}^{SQL})^{2}=(\Delta Q_{n})^{2}+\sigma _{n}^{2}  \label{*}
\end{equation}%
where  $\Delta Q_{n}$ is the variation of $Q_{n}(t)$ and
\begin{equation}
\sigma _{n}^{2}=\sum_{j}(\Delta \xi _{nj})^{2}
\end{equation}%
is just the width of Brownian motion. The first part of the above
equation is just an average over the pure quantum state of the
TLR. Thus we can use the uncertainty principle to determine the
value of SQL. According to the
commutation relation Eq.(\ref{8}), the uncertainty relation means%
\begin{equation}
\Delta \dot{V}_{n}\geq \frac{\hbar \exp (-\gamma t)}{2M_{n}\Delta V_{n}}
\end{equation}%
Then
\begin{equation}
\left( \Delta Q_{n}\right) ^{2}\geq \frac{2\hbar }{M_{n}\omega _{n}^{\prime
2}}e^{-2\gamma t}\left[ \gamma \sin ^{2}\omega _{n}^{\prime }t+\omega
_{n}^{\prime }\sin 2\omega _{n}^{\prime }t\right]
\end{equation}%
It is easy to see that as time $t\rightarrow \infty $, the
standard quantum limit $\left\vert \Delta Q\left( t\right)
\right\vert \rightarrow 0$ (see Fig.1). If the quantum fluctuation
caused by the bath fluctuation $\xi _{nj}\left( t\right) $ is
ignored inappropriately, the standard quantum limit is zero after
a long time evolution! We will show as follows that the quantum
fluctuation contributes a nonzero part as compensation.

\begin{center}
\begin{figure}[tbp]
\includegraphics[width=6cm,height=4cm]{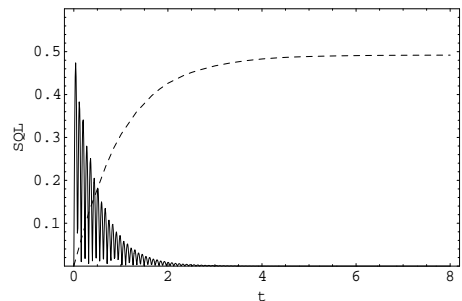}
\caption{The time evolution of the standard quantum limit of the
system in low temperature limit. The solid line shows the
evolution of the SQL of the TLR while the dashed line shows that
of the environment. Here the time is scaled in the unit of
$\frac{1}{\gamma}$ }
\end{figure}
\end{center}

We notice that the second part of Eq. (\ref{*}) is the thermal
average over the reservoir states and the fluctuation $\sigma
_{n}^{2}$ at temperature $T$ is
\begin{equation}
\sigma _{n}^{2}(t)=\sum_{j}\coth \left( \frac{\hbar \omega _{nj}}{2k_{B}T}%
\right) \frac{\hbar \lbrack b_{nj1}^{2}\left( t\right) +\omega
_{nj}^{2}b_{nj2}^{2}\left( t\right) ]}{2m_{nj}\omega _{nj}}.
\end{equation}
where $k_{B}$ is the Boltzman constant and $T$ is the temperature. This is
just the width of the Brownian motion, which characterizes the extent of
fluctuation around the damping path{\bf \ }$Q_{n}$. In the classical limit,
known as the "Ohmic friction" condition, the fluctuation can be evaluated as
\begin{eqnarray}
\sigma _{n}^{2}(t) &\approx &\frac{\hbar \gamma }{\pi M_{n}\omega
_{n}^{\prime }}\int_{0}^{\infty }d\omega \coth \left( \frac{\hbar \omega }{%
2k_{B}T}\right) L\left( \omega \right)  \nonumber \\
&&( 1-2e^{-\frac{\gamma t}{2}}( \gamma \frac{\sin \omega _{n}^{\prime }t}{2}%
+ \omega _{n}^{\prime }\cos \omega _{n}^{\prime }t) \cos \omega t  \nonumber
\\
&&-2\omega \sin \omega _{n}^{\prime }t\sin \omega t )
\end{eqnarray}
where $L(\omega )=\omega/[\left( \omega _{n}^{2}-\omega ^{2}\right)
^{2}+\gamma ^{2}\omega ^{2}]$.

As can be seen from Fig.1, in the low temperature limit $\sigma
_{n}^{2}(t)$ is zero initially, and then approaches its final
equilibrium value in a time interval of the order of $1/\gamma $.
The above results shows the important limit on the standard
quantum limit of TLR caused by the environment fluctuation. It
seemingly depends on the details of the reservoir, but they can be
universally summed up to the observable quantities of the TLR in
certain limit case. When the time is large enough, the total
fluctuation of the whole system approaches (see Fig.2)
\begin{equation}
\left( \Delta V_n\right) _{SQL}^{2}(t\rightarrow\infty )=\frac{\hbar }{2\pi
M_{n}\omega _{n}^{\prime }}\left[ \frac{\pi }{2}+\arctan \left( \frac{\omega
_{n}^{2}}{\gamma \omega _{n}^{\prime }}\right) \right] \,\,
\end{equation}

\begin{center}
\begin{figure}[tbp]
\includegraphics[width=6cm,height=4cm]{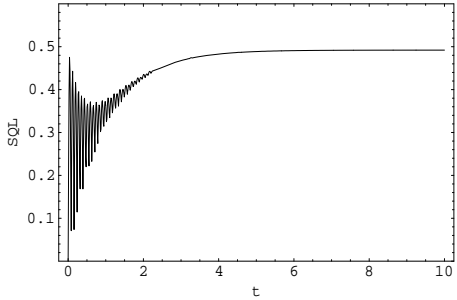}
\caption{The time evolution of the standard quantum limit of the
whole
system at low temperature limit. Again the time is scaled in the unit of $%
\frac{1}{\gamma}$}
\end{figure}
\end{center}

If the damping rate $\gamma $ is much smaller than the frequency of the
oscillator $\omega _{n}$, this width happens to become the same as the width
$\hbar /(2M_{n}\omega _{n})$ of the ground state of the mode, which is just
the SQL of the quadrature amplitudes of the oscillator mode $V_{n}\left(
t\right) $.

To conclude this article let us estimate the above result numerically
according to the parameters given in the experimental proposal in ref \cite%
{Yele1,Yele2}. The eigenfrequency of the TLR mode in resonant with
the Josephson junction qubit is about $10$ GHz while the
dissipation rate is about $6.25\times10^{6}$ Hz. Then
\begin{equation}
\sqrt{\frac{2}{L}}\Delta (V_n)_{SQL}\approx 0.2 \mu V
\end{equation}
where $L$ is the length of the TLR. Thus if the precision of the
experiment can reach this limit, the quantum effect can be
observed. We also note that our exploration in this article
reveals the close relation between the SQL of the open system and
the quantum fluctuation of the environment: starting from the
initial semi-classical state, each mode of TLR experiences a
damping squeezing (see Eqs.(9-11))before it reach the SQL. Once
near the the SQL the quantum fluctuation take place to play as a
quantum noise against the infinite squeezing by quantum
dissipation.

{\it This work is supported by the NSFC and the Knowledge Innovation Program
(KIP) of the Chinese Academy of Sciences. It is also funded by the National
Fundamental Research Program of China with No. 001GB309310. One of the
authors (CPS) thanks Y. B. Dai for useful discussions}.


\begin{references}
\bibitem[a]{email} Electronic address: suncp@itp.ac.cn

\bibitem[b]{www} Internet web site: http:// www.itp.ac.cn/\symbol{126}suncp

\bibitem{Louisell} W.H. Louisell, "Quantum Statistical Properties of
Radiation", John Wiley and Son's, New York, (1990).

\bibitem{dirac} P. A. M. Dirac, Proc. Roy. Soc. A{\bf 133}, (1931) 60.

\bibitem{Yele1} Alexandre Blais, Ren-Shou Huang, Andreas Wallraff, S. M.
Girvin, R. J. Schoelkopf, cond-mat/0402216, 2004.

\bibitem{Yele2} S. M. Girvin, Ren-Shou Huang, Alexandre Blais, Andreas
Wallraff, R. J. Schoelkopf, cond-mat/0310670, 2003.

\bibitem{yele3} R. J. Schoelkopf, A. A. Clerk, S. M. Girvin, K. W. Lehnert,
M. H. Devoret, cond-mat/0210247, 2002. Also in "Quantum Noise in mesoscopic
physics" (Kluwer Ac. Publ., 2003)

\bibitem{Europl} S. Camalet, J. Schriefl, P. Degiovanni, F. Delduc,
cond-mat/0405597, 2004.

\bibitem{landua} L. D. Landau, R. Peierls, Z. Phys. {\bf 69}, 56 (1931).

\bibitem{Bra} V. Braginsky and F. A. Khalili, "Quantum Measurement",
Cambridge, 1992; V. Braginsky and F. A. Khalili, Rev. Mod. Phys, {\bf 68}, 1
(1996).

\bibitem{nano} M. D. LaHaye et al., Science, {\bf 304}, 74 (2004).

\bibitem{ys1} L. H. Yu, C. P. Sun, Phys. Rev. A, {\bf 49}, 592 (1994).

\bibitem{Sun} C. P. Sun, L. H. Yu, Phys. Rev. A, {\bf 51}, 1845 (1995).

\bibitem{Gao} C. P. Sun, Y. B. Gao, Phys. Rev. E, {\bf 49}, 592 (1998)

\bibitem{rem1} To examine it in the two-dimensional case we consider the
Green's theorem
\[
{\oint {\bf I\cdot dl=}}\oint (\nabla \cdot {\bf I)ds}
\]
where ${\bf dl=}(-dy,dx)$ along the direction of tangent. The close path
integral in the left hand side represents the all charges going into the
domain surrounded by the closed path, which is just the sum of all leakages
within this domain. we can consider the integral in the bulk cell $\Delta
x\Delta y$, leading to the Eq.(\ref{cc}).

\bibitem{rma} For the current vector, however, instead of a lossy wave
equationm the modified equation of motion of current vector is
\[
\nabla \left( \nabla \cdot {\bf J}\right) =RG{\bf J}+\left( Rc+lG\right)
\frac{\partial {\bf J}}{\partial t}+lc\frac{\partial ^{2}{\bf J}}{\partial
t^{2}}.
\]
But in one dimension, it can be reduced to the generic lossy wave equation.

\bibitem{ll} L. D. Landau and E. M. Lifshitz, "Quantum Mechanics
(Non-relativistic Theory)", 3rd Edition, Pergamon Press, 1987.

\bibitem{C} P. Caldirola, Nuovo Cimento, {\bf 18}, 393,(1941)

\bibitem{K} E. Kanai, Prog. Theor. Phys. {\bf 3}, 440 (1948)

\bibitem{Dikker} H. Dekker, Phys. Report, {\bf 80}, 1 (1981)

\bibitem{legget} A. O. Caldeira, A. J. Leggett, Ann. Phys. 149, 374 (1983),
and Physica {\bf 121A}, 587 (1983)

\bibitem{WYD} Y. D. Wang, Y. B. Gao and C. P. Sun,
cond-mat/ 0212088, to be published in Euro. Phys. Journal B.

\end{references}
\end{document}